\documentclass[conference]{IEEEtran}
\IEEEoverridecommandlockouts

\usepackage{cite}
\usepackage{amsmath,amssymb,amsfonts}
\usepackage{algorithmic}
\usepackage{graphicx}
\usepackage{textcomp}
\usepackage{xcolor}
\usepackage{enumitem}
\usepackage{algorithm,algorithmic}
\usepackage{subcaption}
\usepackage{hyperref}
\usepackage{array,multirow}
\usepackage{threeparttable}
\usepackage{wasysym}
\usepackage{makecell}
\usepackage{bbold}

\newcommand{\subscript}[2]{$#1 _ #2$}
\def\BibTeX{{\rm B\kern-.05em{\sc i\kern-.025em b}\kern-.08em
    T\kern-.1667em\lower.7ex\hbox{E}\kern-.125emX}}

\makeatletter
\newcommand{\linebreakand}{%
  \end{@IEEEauthorhalign}
  \hfill\mbox{}\par
  \mbox{}\hfill\begin{@IEEEauthorhalign}
}
\makeatother
    
\begin{document}

\title{ When Federated Learning meets Watermarking: A Comprehensive Overview of Techniques for Intellectual Property Protection\\

}

\author{\IEEEauthorblockN{Mohammed Lansari}
\IEEEauthorblockA{\textit{ThereSIS} \\
\textit{Thales SIX GTS}\\
Palaiseau, France \\
mohammed.lansari@thalesgroup.com}
\and
\IEEEauthorblockN{Reda Bellafqira}
\IEEEauthorblockA{\textit{Inserm UMR 1101} \\
\textit{IMT Atlantique}\\
Brest, France \\
reda.bellafqira@imt-atlantique.fr}
\and
\IEEEauthorblockN{Katarzyna Kapusta}
\IEEEauthorblockA{\textit{ThereSIS} \\
\textit{Thales SIX GTS}\\
Palaiseau, France \\
katarzyna.kapusta@thalesgroup.com}
\linebreakand 
\IEEEauthorblockN{Vincent Thouvenot}
\IEEEauthorblockA{\textit{ThereSIS} \\
\textit{Thales SIX GTS}\\
Palaiseau, France \\
vincent.thouvenot@thalesgroup.com}
\and
\IEEEauthorblockN{Olivier Bettan}
\IEEEauthorblockA{\textit{ThereSIS} \\
\textit{Thales SIX GTS}\\
Palaiseau, France \\
olivier.bettan@thalesgroup.com}
\and
\IEEEauthorblockN{Gouenou Coatrieux}
\IEEEauthorblockA{\textit{Inserm UMR 1101} \\
\textit{IMT Atlantique}\\
Brest, France \\
gouenou.coatrieux@imt-atlantique.fr}
}

\author{
    \IEEEauthorblockN{
        Mohammed Lansari\IEEEauthorrefmark{1}\IEEEauthorrefmark{2}\IEEEauthorrefmark{3}, Reda Bellafqira\IEEEauthorrefmark{1}\IEEEauthorrefmark{3}, Katarzyna Kapusta\IEEEauthorrefmark{2}, \\Vincent Thouvenot\IEEEauthorrefmark{2}, Olivier Bettan\IEEEauthorrefmark{2}, Gouenou Coatrieux\IEEEauthorrefmark{1}
    } \\
    \IEEEauthorblockA{\IEEEauthorrefmark{1} Inserm UMR 1101, IMT Atlantique \\ Brest, France \\ name.surname@imt-atlantique.fr} 
    \\
    \IEEEauthorblockA{\IEEEauthorrefmark{2} ThereSIS, Thales SIX GTS \\ Palaiseau, France \\ name.surname@thalesgroup.com}
}

\maketitle

\def\thefootnote{$\ddagger$}\footnotetext{Equal contributions.}
\begin{abstract}
Federated Learning (FL) is a technique that allows multiple participants to collaboratively train a Deep Neural Network (DNN) without the need of centralizing their data. Among other advantages, it comes with privacy-preserving properties making it attractive for application in sensitive contexts, such as health care or the military. Although the data are not explicitly exchanged, the training procedure requires sharing information about participants' models. This makes the individual models vulnerable to theft or unauthorized distribution by malicious actors. To address the issue of ownership rights protection in the context of Machine Learning (ML), DNN Watermarking methods have been developed during the last five years. Most existing works have focused on watermarking in a centralized manner, but only a few methods have been designed for FL and its unique constraints. In this paper, we provide an overview of recent advancements in Federated Learning watermarking, shedding light on the new challenges and opportunities that arise in this field.

\end{abstract}

\begin{IEEEkeywords}
DNN Watermarking, Federated Learning, Intellectual Property, Security.
\end{IEEEkeywords}

\section{Introduction}
 In the fast-paced digital era, machine learning (ML) has emerged as a revolutionary technology that drives innovation across various industries. Its ability to analyze vast amounts of data and make accurate predictions has opened up a world of possibilities. From enabling personalized recommendations in e-commerce \cite{singh2020commerce} and improving medical diagnoses \cite{conze2023time} to enhancing autonomous vehicles \cite{mallozzi2019autonomous}, the applications of machine learning are both diverse and impactful.

The effectiveness of machine learning models depends on the availability of large, high-quality datasets. Traditionally, the success of ML algorithms relied on centralized data storage and processing, where organizations and institutions accumulated vast amounts of information to train their models effectively. Unfortunately, this approach increases the risk of privacy leakage, as sensitive information about individuals, customers, or proprietary processes may inadvertently be exposed to unauthorized entities. This presents a roadblock to the advancement of machine learning, as data holders are understandably hesitant to compromise their users' privacy and organizational security. Moreover, safeguarding the privacy and security of sensitive data is not just an ethical obligation but a legal necessity as highlighted by various data protection regulations around the world, such as the General Data Protection Regulation (GDPR) \cite{regulation2018general} in EU, California Consumer Privacy Act (CCPA) \cite{piper2019data} in US, and Data Security Law (DSL) \cite{chen2021understanding} in China.

Addressing these challenges, Federated Learning (FL) \cite{mcmahan2017communication} emerges as a promising solution. FL enables multiple data owners or distributed agents to collectively train a global model without directly sharing their private data. Incorporating privacy-preserving techniques like Homomorphic Encryption \cite{benaissa2021tenseal}, Multi-Party Computation \cite{gehlhar2023safefl}, Private set intersection (PSI) \cite{chen2017fast}, Differential Privacy \cite{dwork2008differential, wei2023personalized}, or hardware-based solutions such as trusted execution environment (TEE) \cite{zheng2023secure}, FL ensures data privacy and security while enabling collaborative model training. The significance of federated learning lies not only in its ability to alleviate privacy concerns but also in its potential to unlock the untapped value of distributed data. By enabling cooperation among organizations and institutions, federated learning allows for more comprehensive and representative models, ultimately leading to more accurate predictions and valuable insights \cite{xu2023federated}.

Another critical aspect to take into account is the protection of the intellectual property of the trained model, aimed at preventing theft, plagiarism, and unauthorized usage by external parties. In response to this concern, model watermarking has been introduced since 2017 as a solution to embed a unique watermark or fingerprint into the model. This measure effectively safeguards the model's intellectual property and enables tracing the source in case of any illicit model leakage. However, the majority of model watermarking methods address the problem of local or centralized training. The collaborative setting of Federated Learning raises new challenges in the context of ownership protection. How to identify that a participant contributed to the training of a model resulting from Federated Learning? How to be sure that the final model will not be misused by the aggregation party that coordinates the Federated Learning? These issues may slow down parties from joining the learning consortium.  Previous work by Yang \textit{et al.}  \cite{yang2023federated} has conducted a survey on federated watermarking solutions, primarily focusing on security issues within Federated Learning. However, the paper only presents two watermarking algorithms (Waffle \cite{tekgul2021WAFFLE} and FEDIPR \cite{li2022fedipr}) specifically designed for the context of FL.

\subsection{Contributions}
Our paper makes the following main contributions:

\begin{itemize}
\item We emphasize the foundational principle of secure federated learning and stress the importance of comprehensive security guarantees that align with the trust levels of the participants. This includes addressing various aspects such as training on non Independently and Identically Distributed (non-I.I.D) data and ensuring robustness against poisoning and backdooring attacks.
\item We conduct a comprehensive examination of the challenges concerning model watermarking in the context of Federated Learning.
\item We expand upon the work presented in \cite{yang2023federated} by providing an in-depth analysis of nine currently available DNN watermarking schemes in the FL context, along with their limitations.
\item We bring attention to unresolved issues that necessitate further exploration.
\end{itemize}

\subsection{Organization}
The rest of the paper is organized as follows. Section \ref{background} describes background information on Federated Learning and DNN watermarking concept. Section \ref{WMforFL} presents our definition of watermarking for Federated Learning followed by an overview of the six existing works combining FL and DNN. In Section \ref{discussion}, we analyze then the different elements that have to be taken into consideration while designing a watermarking scheme for a collaborative training setting. We identify the unsolved challenges and thus give an insight into possible future works. Section \ref{sec:conclusion} concludes the paper.

\section{Background}
\label{background}
In this section, we remind the definition of Federated Learning and give a brief overview of its different existing settings. Then we introduce DNN watermarking.
\subsection{Federated Learning}
\label{section:fl}
Federated Learning is a machine learning setting where $K \in \mathbb{N}^*$ entities called clients collaborate to train a global model $M_G$ while keeping the training data $D_C = \{ D_{C_i} \}_{i=1}^K$ decentralized \cite{kairouz2021advances}. 

The most popular framework is called Client-Server FL or scatter $\&$ gather framework. Here's a high-level overview of this setting:
\begin{enumerate}
    \item Setup:
    \begin{itemize}
        \item Central Server: A central server manages the overall training process and initiates model updates.
        \item Clients: These are the individual devices, such as smartphones, IoT devices, or computers, that participate in the training process. Each edge device has its local dataset that cannot be shared with the central server due to privacy concerns.
    \end{itemize}
    \item Initialization:
    \begin{itemize}
        \item Initially, the central server initiates a global model $M_G$ (e.g. with random parameters) and distributes it to a subset of clients selected randomly at each round.
    \end{itemize}
    \item Local Training:
     \begin{itemize}
         \item Each client trains the global model $M_G$ on its local dataset using its computational resources. The training is typically performed using gradient descent or a similar optimization algorithm.
     \end{itemize}    
    \item Model Update:
    \begin{itemize}
        \item After the local training is complete, the clients generate a model update (typically gradients) based on the locally processed data.
    \end{itemize}
    \item Aggregation:
    \begin{itemize}
        \item The clients send their model updates back to the central server without sharing their raw data.
        \item The central server aggregates all the received model updates to create a refined global model. This is usually done by averaging the model updates (see e.g.\cite{brendan2016communication}).
    \end{itemize}
    \item Iterative Process:
    \begin{itemize}
        \item Steps 3 to 5 are repeated for multiple rounds or epochs, allowing the global model to improve over time by leveraging knowledge from various clients.
    \end{itemize}
    \item Centralized Model Deployment:
    \begin{itemize}
        \item Once the federated training process is complete, the final global model can be deployed from the central server to all clients for local inference.
    \end{itemize}

\end{enumerate}

Note that, the presence of the server is not mandatory to perform FL. In a decentralized setting also known as a cyclic framework, clients can perform FL without the supervision of a server. BrainTorrent \cite{roy2019braintorrent} proposes a server-less and peer-to-peer Federated framework. In this solution, a random client is selected to be the aggregator. Then, it checks if other clients have an updated version of the model. If yes, they send it to the aggregator who performs an averaging of the model weights/updates. Then it updates its own model with the previous result.

\begin{table*}[t]
  \centering
\begin{tabular}{|c|c|c|c|}
\hline
\thead{FL frameworks} & \thead{Developed by} & \thead{Purpose} & \thead{Security protocols provided}  \\ 
\hline
Fed-BioMed \cite{cremonesi2023fed} & INRIA & Research  & DP, HE \\
\hline
TensorFlowFederated \cite{tff}  &  Google & Research  &  DP \\
\hline
PySyft \cite{ziller2021pysyft} & OpenMined & Research  & MPC, DP, HE, PSI  \\
\hline
Flower \cite{beutel2020flower} & Flower Labs GmbH & Industrial & DP  \\
\hline
FATE \cite{fate} & WeBank & Industrial & HE, DP, MPC  \\
\hline
OpenFL \cite{reina2021openfl} & Intel & Industrial  &  TEE \\
\hline
IBM Federated Learning \cite{ludwig2020ibm} & IBM&  Industrial  & DP, MPC  \\ 
\hline
NvFlare \cite{roth2022nvidia} & Nvidia & Industrial  & HE, DP, PSI \\ 
\hline
Clara \cite{clara} & Nvidia & Industrial & DP, HE, TEE \\
\hline

\end{tabular}
\caption{Examples of Open-source Federated Learning (FL) frameworks
}
\label{tab:fl_frameworks}
\end{table*}

Another type of decentralized learning is Split-Learning \cite{vepakomma2018split} which consists of splitting the DNN model between the server and the clients. Many configurations are possible but the most common for client privacy is the U-shaped configuration. The aim is that each client has the first and last layers of the DNN model and the server has the rest of the layers. In such a way, clients perform the forward/backward pass keeping their data (inputs and labels) private, and send/receive only the activation maps/gradients, respectively,  to update the whole model.

The FL is also defined by the data features partitioning, which can be categorized into three types: horizontal, vertical, and hybrid FL. In the case of horizontal FL, all clients share the same set of features, but their sample data may differ. In contrast, vertical FL assumes that all clients possess the same sample data but have distinct features. Lastly, hybrid FL involves clients with varying samples and features across the board. 

Ideally, the data should be Independently and Identically Distributed (I.I.D) for each client. However, in real-world FL scenarios, data distribution and amount differ between clients since they collect and use their own data. This leads to a non-I.I.D data partition, which can result in significant performance loss \cite{zhao2018federated}. To address this issue, several aggregation functions have been proposed, such as FedProx \cite{li2020federated} and SCAFFOLD \cite{karimireddy2020scaffold}. These methods aim to improve the performance of federated learning despite the non-I.I.D data distribution across clients.

The aim of federated learning is to ensure the security and privacy of clients' data while achieving a model's performance equivalent to centralized training. However, this objective can be compromised if the server or certain clients act maliciously. In real scenarios, the server may be considered an honest but curious entity, meaning it will abide by the FL protocol regarding client selection and global model aggregation/distribution, but it will attempt to infer information about the client data. This situation commonly arises in classical federated learning setups, where only model parameters/gradients are shared. The server can potentially employ attacks like the inversion attack, which reconstructs the data used for learning from the gradients shared by the clients. To counteract this type of attack, various countermeasures have been developed, one of which is homomorphic encryption (HE) \cite{benaissa2021tenseal}. Homomorphic encryption enables linear operations to be performed on data from their encrypted form (without having access to the secret key), preventing the server from conducting inversion attacks on encrypted gradients. However, using HE has some drawbacks, including high computational and communication complexity, and it also restricts the range of aggregation methods, as it only allows linear operations to be performed on the encrypted data.

Another approach to address privacy concerns is Differential Privacy, which combats inversion attacks by introducing noise to the updated gradients \cite{dwork2008differential, wei2023personalized}. Nevertheless, this technique faces a limitation due to the trade-off between utility and privacy. Alternatively, Trusted Execution Environments (TEEs) \cite{mo2020darknetz,zheng2023secure} present another solution, where a trusted third party is utilized to provide guarantees for code and data confidentiality and integrity. Multi-party computation (MPC) has also been implemented to achieve secure model aggregation in federated learning, as demonstrated in \cite{kanagavelu2020two, mo2020darknetz, zheng2023secure}. However, similar to encryption-based solutions, MPC-based approaches are susceptible to efficiency issues. In addition to reinforcing the confidentiality of the aggregated data, MPC - and more precisely its private set intersection protocol \cite{chen2017fast} - can be applied to identify the intersection of feature spaces between clients in vertically partitioned data.

In scenarios where a group of clients consists of malicious users, unlike the server, these clients have access to their local data and models. Leveraging this knowledge, they can employ various attacks to undermine the model's performance. Some of these attacks include poisoning attacks \cite{fang2020local, shi2022challenges}, which aim to introduce malicious data to corrupt the model's training or attacks that cause the model to misclassify data with specific patterns while maintaining its performance on the primary task, known as backdooring attacks \cite{huang2020dynamic, xie2019dba}. These malicious actions pose significant challenges to the security and integrity of the federated learning process.

To defend against poisoning and backdooring attacks in the context of federated learning, two classes of approaches have been proposed. The first class involves developing robust aggregation techniques, such as Krum aggregation \cite{blanchard2017machine} or median and trimmed mean aggregation \cite{yin2018byzantine}, which are designed to identify and remove abnormal local models contributed by potentially malicious clients. These techniques help improve the overall model's resilience to attacks. The second class of defenses relies on the use of anomaly detection techniques implemented by the server at each round of federated learning. These anomaly detection methods enable the server to identify and filter out abnormal client updates before performing the model aggregation process \cite{anass2022poisoning, gu2021detecting, li2020learning}. By doing so, the server can mitigate the impact of potential malicious clients and enhance the security of the federated learning system. For more comprehensive insights into the threats and defenses related to federated learning, interested readers can refer to the work cited in \cite{shi2022challenges}.

To summarize, to establish a secure federated learning (FL) environment, the first step is to conduct a security analysis of the entities engaged in FL and assess their level of trust. This analysis helps identify potential threats and risks specific to the scenario being considered. Once the security analysis is complete, the next step involves integrating the security tools discussed earlier to design a robust FL model that performs effectively while ensuring data and model privacy. In Table \ref{tab:fl_frameworks}, we present a list of secure federated learning frameworks, along with the corresponding security tools that are employed to provide the necessary security measures. Furthermore, when developing an efficient and practical watermarking solution for federated learning, it is crucial to consider all the aforementioned security requirements. We elaborate on how these constraints can be accommodated in the context of federated learning watermarking in Section \ref{discussion}.

\subsection{DNN Watermarking}
\label{subsec:dnnwatermarking}

\begin{table*}[t]
  \centering
\begin{tabular}{|l|l|}
\hline
\textbf{Fidelity} & The watermarked model needs to have performances as close as possible compared to the model without watermark \\ \hline
\textbf{Capacity} &  The capacity of a technique to embed multiple watermarks \\ \hline
\textbf{Reliability} &  Demonstrate a low false negative rate, enabling legitimate users to accurately identify their intellectual property with a high level of certainty.\\ \hline
\textbf{Integrity} &   Demonstrate a low false positive rate, preventing unjustly accusing honest parties with similar models of intellectual property theft. \\ \hline
\textbf{Generality} & The capacity of a watermarking technique to be applied independently of the architecture of the model \\ \hline
\textbf{Efficiency} & The performance cost generated by the embedding and verification process of the watermarking \\ \hline
\textbf{Robustness} & The capacity to resist against attacks aiming at removing the watermark \\ \hline
\textbf{Secrecy} & The watermark should be secret and undetectable \\ \hline
\end{tabular}
\caption{DNN Watermarking requirements and properties}
\label{requirements}
\end{table*}
\subsubsection{Requirements}

DNN watermarking is a promising solution for ownership protection of ML models \cite{xue2021dnn, lukas2022sok, li2021survey, fkirin2022copyright, boenisch2021systematic, sun2023deep}. Inspired by multimedia and database watermarking \cite{ bouslimi2016data, ernawan2023recent, pavan2023overview, niyitegeka2018dynamic, hu2023reversible}, it consists in introducing a secret change into the model parameters or behavior during its training, in order to enable its identification in the future. As multimedia content watermarking, DNN watermarking must respect some requirements to be effective for IP protection. Table \ref{requirements} summarizes these requirements. 

The watermark must be secret (Secrecy). This requirement refers to the fact that any person who analyzes the model is not able to detect if the latter is watermarked. White-Box techniques can change the distribution of the parameters and then differs from a non-watermarked model. In addition, a watermarking technique must also preserve the model's performance on the main task (Fidelity). If the watermarking embedding process returns a model that has an accuracy up to a defined $\epsilon$ compared to a non-watermarked model, then the watermarking technique is not efficient. Reliability ensures a low false negative rate for the owner during IP verification, while Integrity aims to prevent false positive claims by other parties. The watermarking algorithm should be independent of the model (Generality) while providing a large insertion Capacity, which can be either zero-bit, indicating only the presence of a watermark, or multi-bit, allowing the encoding of multiple bits of information.

Furthermore, the watermark must exhibit Robustness against attacks aimed at removing or detecting it. A removal attack is considered effective if it maintains a high test accuracy while eliminating the watermark. It is efficient if the resources required for the attack, such as runtime, are relatively small compared to retraining the model from scratch. Some common attacks include:
\begin{itemize}
    \item Pruning Attack: Setting the less useful weights of the model to zero.
    \item Fine-Tuning Attack: Re-training the model and updating its weights without decreasing accuracy.
    \item Overwriting Attack: Embedding a new watermark to replace the original one.
    \item Wang and Kerschbaum Attack: For static white-box watermarking algorithms, it alters the weight distribution of the watermarked model, relying on visual inspection.
    \item  Property Inference Attack: Training a discriminating model to distinguish watermarked from non-watermarked models, thereby detecting if a protected model is no longer watermarked.
    \item Another attack is the Ambiguity Attack, which forges a new watermark on a model, making it challenging for external entities, like legal authorities, to determine the legitimate watermark owner. This ambiguity prevents the legitimate owner from claiming the copyright of the intellectual property.

\end{itemize}

\subsubsection{Related works}
DNN Watermarking can be distinguished into two types of techniques: White-Box  \cite{song2017machine, uchida2017embedding, feng2020watermarking, li2020spread, tartaglione2021delving, chen2019deepmarks, wang2020watermarking, rouhani2019deepsigns, wang2020watermarking, fan2019rethinking, li2021secure,  bellafqira2022diction, kuribayashi2023white, lv2023robustness, rodriguez2023towards} and Black-Box \cite{rouhani2019deepsigns, chen2019blackmarks, vybornova2022method, zhang2018protecting, adi2018turning, guo2018watermarking, le2020adversarial, namba2019robust, li2019prove, kapusta2020watermarking, lounici2022blindspot, kallas2022rose, qiao2023novel, kallas2023mixer, hua2023unambiguous} watermarking. Each technique is defined by the type of access to the model parameters during the verification process. 

In the White-Box setting, we assume that the owner will have full access to the model (architecture, parameters, activation maps...). In this way, to insert a watermark into a DNN, the owner will hide a piece of information $b$ in the form of a binary string or an image (e.g. Quick Response code) into the model's parameters \cite{uchida2017embedding, wang2019riga, rodriguez2023towards}, activation maps \cite{rouhani2019deepsigns, li2021secure, bellafqira2022diction} or by adding a passport layer \cite{fan2019rethinking, zhang2020passport}. As formulated in \cite{bellafqira2022diction}, a white box watermarking scheme is defined as follows:

\begin{enumerate}
    \item Initially, a target model $M$ is considered, and a features extraction function $Ext(M,K_{ext})$ is applied with a secret key $K_{ext}$. The features obtained can be a subset of the model weights, where $K_{ext}$ indicates the indices of the selected weights. Alternatively, the features can be model activation maps for specific input data secretly chosen from a trigger set. These features are then utilized for watermark insertion and extraction.
    \item The embedding of a watermark message $b$ involves regularizing $M$ using a specific regularization term $E_{wat}$. This regularization term ensures that the projection function $Proj(.,K_{Proj})$ applied to the selected features encodes the watermark $b$ in a predetermined watermark space, which depends on the secret key $K_{Proj}$. The goal is to achieve the following after training:
\begin{equation}
Proj(Ext(M^{wat}, K_{ext}), K_{Proj}) = b
\end{equation}
where $M^{wat}$ is the watermarked version of the target model $M$.
 To achieve this, the watermarking regularization term $E_{wat}$ relies on a distance measure $d$ defined in the watermark space. For example, in the case of a binary watermark with a binary string of length $l$, i.e., $b \in \{0,1\}^l$, the distance measure could be the Hamming distance, Hinge distance, or Cross-Entropy. The regularization term is formulated as:
\begin{equation}\label{eq:loss_wat}
E_{wat} = d(Proj(Ext(M^{wat}, K_{ext}), K_{Proj}), b)
\end{equation}

 To preserve the accuracy of the target model, the watermarked model $M^{wat}$ is usually derived from $M$ through a fine-tuning operation parameterized with the following loss function:
\begin{equation}
\label{eq:loss_global_wat}
E = {E}_{0}(X_{Train}, Y_{Train}) + \lambda E_{wat}
\end{equation}
where $E_{0}(X_{Train}, Y_{Train})$ represents the original loss function of the network, which is essential to ensure good performance in the classification task. $E_{wat}$ is the regularization term added to facilitate proper watermark extraction, and $\lambda$ is a parameter that adjusts the trade-off between the original loss term and the regularization term.

\item The watermark retrieval process is relatively straightforward. It involves using both the features extraction function $Ext(.,K_{ext})$ and the projection function $Proj(.,K_{Proj})$ as follows:
\begin{equation}
b^{ext} = Proj(Ext(M^{wat},K_{ext}),K_{Proj})
\end{equation}
where $b^{ext}$ is the extracted message from the watermarked model $M^{wat}$.
\end{enumerate}

For example, in the watermarking scheme introduced by Uchida \textit{et al.} \cite{uchida2017embedding}, the feature extraction function $Ext(., K_{ext})$ involves computing the mean value of secretly selected filter weights $\mathbf{w}$, where $K_{ext}$ represents the index of the chosen convolutional layer. The projection function $Proj(., K_{Proj})$ in \cite{uchida2017embedding} is designed to insert a watermark $b \in \{0, 1\}^l$, where $l$ is the length of the watermark, and it is defined as follows:

\begin{equation}
\label{eq:uchida_proj}
Proj(\mathbf{w}, K_{Proj}) = \sigma(\mathbf{w} K_{Proj}) \in \{0, 1\}^l
\end{equation}

Here, $K_{Proj}$ represents a secret random matrix of size $(|w|,l)$, and $\sigma (.)$ is the Sigmoid function:
\begin{equation*}
\sigma(x)=\frac{1}{1+e^{-x}}
\end{equation*}

Uchida \textit{et al.} \cite{uchida2017embedding} use binary cross entropy as the distance measure $d$ for the watermarking regularization $E_{wat}$, given by:
\begin{equation}\label{regulization:cross_entropy}
d(b,y) = -\sum_{j=1}^{l}(b_{j}\log(y_{j})+(1-b_{j})\log(1-y_{j}))
\end{equation}

With the information on $d$, $Proj(., K_{Proj})$, and $Ext(., K_{ext})$, the loss $E_{wat}$ for watermarking a target model $M$ can be computed using \eqref{eq:loss_wat}.

Another example of a watermarking scheme proposed in \cite{yang2023fedzkp},  where the feature extraction function $Ext(., K_{ext})$ consist of the scaling parameters of the Batch Normalization (BN) weights $W_\gamma = (\gamma^1, \gamma^2,..., \gamma^l)$ (defined in Eq. \eqref{eq:bn} ) with $l$ channels is chosen according to the secret position parameter $K_{ext}$.
\begin{equation}
    \label{eq:bn}
    y^{i} = \gamma^i*x^i + \beta^i 
\end{equation}

where $\gamma^i$  and $\beta^i$  are the scaling and bias parameters in channel $i$ for BN layer respectively, $x^i$ is the input of the BN layer. The projection function $Proj(., K_{Proj})$ in \cite{yang2023fedzkp} is designed to insert a watermark $b \in \{0, 1\}^l$, where $l$ is the length of the watermark, and it is defined as follows:

\begin{equation}
\label{eq:uchida_proj}
Proj(W_\gamma, K_{Proj}) = Sgn(W_\gamma K_{Proj}) \in \{0, 1\}^l
\end{equation}

where $K_{Proj}$ is smiliar to Uchida \textit{et al.} scheme \cite{uchida2017embedding}, which  represents a secret random matrix of size $(|w|,l)$, and $Sgn (.)$ is the sign function:
\begin{equation}
Sgn(x)=\left\{
    \begin{array}{ll}
        1 & \mbox{if } x>0 \\
        0 & \mbox{Otherwise.}
    \end{array}
\right.
\end{equation}

The Hinge Loss is used as the distance measure $d$ for the watermarking regularization $E_{wat}$, given by:
\begin{equation}\label{regulization:hinge_loss}
d(b,y) = -\sum_{j=1}^{l} \max(\mu - b_i y_i, 0) 
\end{equation}
Here, $\mu$ represents the parameter of the Hinge Loss as defined in \cite{rosasco2004loss}. Similar to the scheme proposed by Uchida \textit{et al.} in \cite{uchida2017embedding}, and utilizing the information about $d$, $Proj(., K_{Proj})$, and $Ext(., K_{ext})$, the watermarking loss $E_{wat}$ for the purpose of watermarking a target model $M$ can be computed using the equation \eqref{eq:loss_wat}.
  
\begin{figure*}
    \centering
  \includegraphics[width=0.8\textwidth]{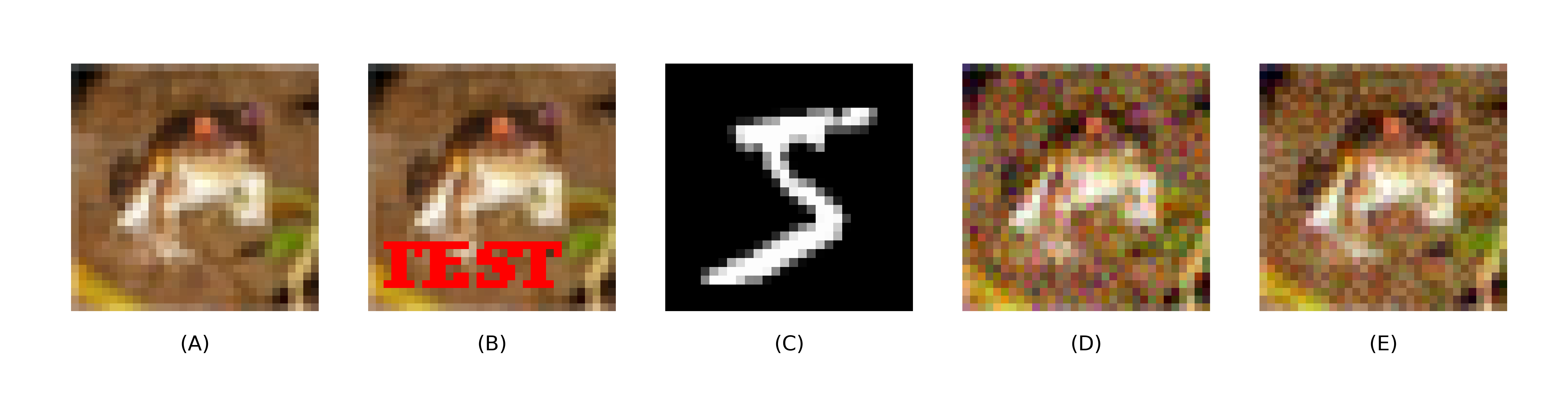}
  \caption{Sample from possible trigger sets : (A) Original (B) Content (C) Unrelated (D) Noise (E) Adversarial}
  \label{fig:triggerset}
\end{figure*}

On the other hand, the Black-Box setting assumes that the owner can perform the verification process only through an API: he can interact with the model only by giving inputs and receiving associated predictions. Knowing that the owner watermarks the model by changing its behavior. The common technique consists training of the model using a trigger set $T = (X_i,Y_i)_{i=1}$, which is composed of crafted inputs $X_i$ with their associated outputs $Y_i$ \cite{adi2018turning}. During each epoch, instead of giving a batch only from the train set to the model, we give a concatenation between a batch from the train set and the trigger set.

For example, Zhang \textit{et al.} \cite{zhang2018protecting} propose to use the same technique but with different types of inputs. They try three different types of images for the trigger set that are illustrated in Figure \ref{fig:triggerset} :
\begin{enumerate}
    \item Content Watermarking: Adding meaningful content to images from the train set. The model should be triggered by the content and returns the associated fixed label. In our example, the text "TEST" is used to trigger the model.
    \item Unrelated Watermarking: Images that are irrelevant from the main task of the model. Each image has an associated label (like in \cite{adi2018turning}) or each sample can have its specific output. In our example, some images from the MNIST dataset are used to trigger the model.
    \item Noise Watermarking: Adding a specific noise to images from the train set. Then the model classifies any images with this specific noise as a predefined label. In our example, we add a small Gaussian noise to trigger the model.
\end{enumerate}

The trigger set can also be built using adversarial examples. Authors of \cite{le2020adversarial} proposed to use a trigger set composed of two adversarial examples : 

\begin{enumerate}
    \item True adversaries: Samples that are miss-classified by the model while being close to being well classified.
    \item False adversaries: Well-classified samples from which we add an adversarial perturbation without changing their classification.
\end{enumerate}

Then the model is trained to well classify the true adversaries with their true associated labels and the false adversaries with their labels. 
In Figure \ref{fig:triggerset} we use Fast Gradient Sign Method \cite{goodfellow2014explaining} to generate a possible False adversary.

To evaluate the performance of BlackBox watermarking embedding, we assess the accuracy of the model's output on the trigger set $T$ and their labels as follows:

\begin{equation}
    acc = \frac{1}{|T|} \sum_{(x,y) \in T} \mathbb{1}_{M^{wat}(x) = y}
\end{equation}

\section{Watermarking for Federated Learning}
\label{WMforFL}

In this section, we introduce and define what is watermarking for Federated Learning including the different possible scenarios. Then, we formulate requirements for watermarking deployed in a Federated Learning context. Finally, we analyze the nine state-of-the-art methods.

\subsection{Definition}
\label{def:wfl}
In the context of centralized DNN watermarking, the goal is to simply prove the model's ownership subsequent to its training and release. In FL, ownership rights protection becomes a more complex problem due to the presence of multiple participants and multiple exchanges between them that have to be taken into account during the threat model formulation. To illustrate this issue, the authors in  \cite{tekgul2021WAFFLE} show that the existing methods can naively be applied to FL in two manners: The first one consists in watermarking the model after the training is completed. For example, by using fine-tuning to embed the watermark into the trained model. Without taking the fidelity requirement into account, any participant that receives the model (client or server) can steal the DNN before the last round. The second way is to embed the watermark before starting the training process. Even if the watermark will resist during the first rounds, it will be removed after several aggregation rounds. Thus, it is important to design a specific watermarking technique for FL that will be persistent from the first round until the model deployment.

\begin{figure*}[ht]
    \centering
    \includegraphics[width=0.90\textwidth]{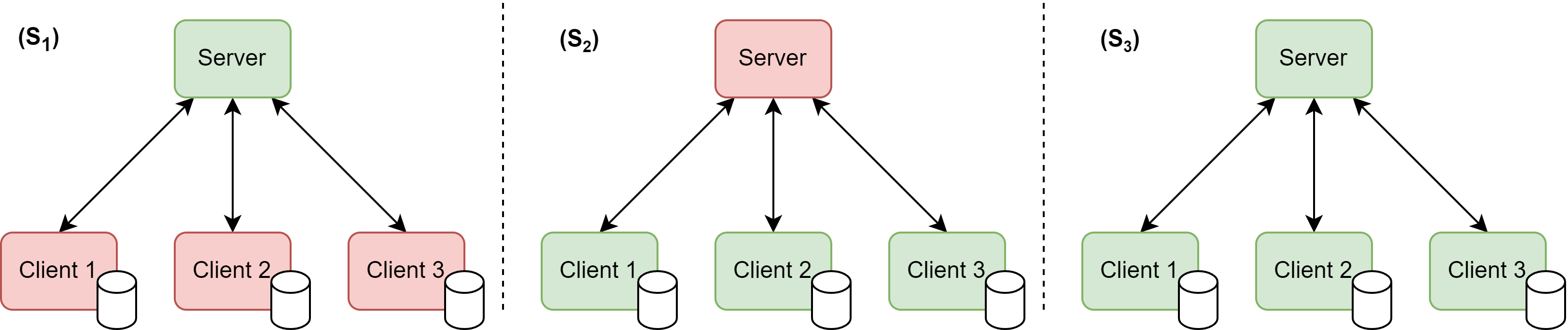}
    \caption{An example of each possible scenario of watermarking in FL with one server and three clients. Green rectangles are the participants who follow the same watermarking procedure together. Red rectangles are those who are not enrolled in the watermarking procedure.}
    \label{fig:scenarios}
\end{figure*}

We define Watermarking for Federated Learning as the process for a participant or multiple participants to insert a watermark into the shared model. Following the client-server FL framework, the first question is to determine which part of the federation can watermark the model. Is the server more to be trusted since it manages the federation? Or the clients since their data are used? During this study, we distinguish three watermarking scenarios FL context, which we illustrate in Figure \ref{fig:scenarios} according to who is watermarking the model.

\begin{enumerate}[label=(\subscript{\textbf{S}}{\textbf{{\arabic*}}})]
    \item \textbf{Server}: The server is in charge of watermarking the global model.  
    \item \textbf{Clients}: One or multiple clients watermark their updates in order to watermark the global model.
    \item \textbf{Server and Clients}: The server and the clients collaborate to watermark the global model together.

\end{enumerate}

All watermarking requirements defined in Table \ref{requirements} are also true in the federated context. However, due to the new constraints and the extension to several participants, we can add precision to five of them:

\begin{enumerate}[]
    \item \textbf{Capacity:} When multiple clients want to insert their own message $b_{C_i}$, the watermarking technique needs to avoid possible conflict between the different inserted $b_{C_i}$. The number of bits needs to be then enough.  
    \item \textbf{Generality:} In a real FL scenario, many additional mechanisms are added for security and privacy such as robust aggregation functions (Section \ref{aggregation}) or Differential Privacy \cite{abadi2016deep} (Section \ref{differential}). The watermarking technique must be applied independently to these mechanisms.
    \item \textbf{Efficiency:} The cost generated by the embedding process is more crucial in FL. For example, in a cross-device architecture, clients have low computation power and they cannot perform many operations. The watermarking techniques must take this parameter into account.
    \item \textbf{Secrecy:} If all parties are not enrolled in the watermarking process, the watermark should not be detected. In particular, if one or some clients are trying to watermark the global model, their updates need to be similar to benign updates. Otherwise, the server can use a defensive aggregation function to cancel the FL watermarking process (as described in \ref{section:fl}).
    \item \textbf{Robustness:} Since the model can be redistributed for any clients or the server, the watermark must track who is the traitor. Traitor tracing is the fact that each actor of the federation has a unique watermark that can be used to uniquely identify the owner in addition to a global watermark.
\end{enumerate}

\subsection{Related works}

\begin{table*}
    \centering
    \captionsetup{justification=centering}
    \begin{tabular}{|c|c|c|c|c|c|c|c|c|c|}
       \hline
       \multirow{2}{*}{\thead{Existing Works}}           & \multirow{2}{*}{ \thead{Verification}}      & \multirow{2}{*}{\thead{Watermarks \\ embedding}} & \multicolumn{4}{c|}{\thead{Security Tools Compatibility}} &  \multirow{2}{*}{\thead{Clients \\ Selection }} & \multicolumn{2}{c|}{\thead{Poisoning Defense}}  \\
       \cline{4-7} \cline{9-10}
                                                    &                       &                    & DP & HE  & MPC & TEE &  & Aggregation & Anomaly Detector \\
        \hline
        WAFFLE \cite{tekgul2021WAFFLE}             &  Black-Box             & Server            &    \LEFTcircle    &  -  &  \LEFTcircle  & \LEFTcircle & \LEFTcircle  & \LEFTcircle  & \LEFTcircle \\
        \hline
        FedIPR \cite{li2022fedipr}                 & White-Box \& Black-Box &         Client(s)   &    \CIRCLE    &  \LEFTcircle   &    \LEFTcircle    & \LEFTcircle & \CIRCLE  & \CIRCLE & \LEFTcircle \\
        \hline
         FedTracker \cite{shao2022fedtracker}      & White-Box \& Black-Box & Server           &   \LEFTcircle  &  -  &   \LEFTcircle   &   \LEFTcircle   &  \LEFTcircle  &  \LEFTcircle &  \LEFTcircle \\
        \hline
        Liu et al. \cite{liu2021secure}            & Black-Box              &         Client &   \LEFTcircle  &  \CIRCLE &   \LEFTcircle   &  \LEFTcircle  & \LEFTcircle  & - & - \\
        \hline
        FedCIP \cite{liang2023fedcip}              &  White-Box             &         Client(s)   &   \LEFTcircle      &  \LEFTcircle  &    \LEFTcircle    &  \LEFTcircle   & \CIRCLE  & \LEFTcircle & \LEFTcircle \\ 
        \hline
        FedRight  \cite{chen2023fedright}          &  White-Box             & Server         &   \LEFTcircle     &  -   &   \LEFTcircle   &   \LEFTcircle  &  \LEFTcircle &  \LEFTcircle &  \LEFTcircle \\
        \hline
        Yang et al.  \cite{yang2022watermarking}          & Black-Box     &  Client        &    \LEFTcircle  &  \CIRCLE   &  \LEFTcircle   &  \LEFTcircle   &  \LEFTcircle &  - & -  \\
        \hline
        FedZKP \cite{liang2023fedcip}              &  White-Box             & Client(s)            &    \LEFTcircle     &   \LEFTcircle &    \LEFTcircle     &   \LEFTcircle   &  \LEFTcircle  & \LEFTcircle  & \LEFTcircle  \\
        \hline
        Merkle-Sign  \cite{li2021towards}          & White-Box \&  Black-Box             & Server            &    \LEFTcircle   &   -  &   \LEFTcircle      & \LEFTcircle & \LEFTcircle  & \LEFTcircle  & \LEFTcircle  \\
        \hline
    \end{tabular}

    \caption{FL watermarking taxonomy, categorized according to the access required to the verifier (white or black box), the entity that embeds the watermark (server, one or multiple clients), and their compatibility with the cryptographic security tools \\ \CIRCLE: Compatible and tested, \LEFTcircle: Compatible but not tested, -: Not compatible}
    \label{tab:fl_watermarking}
\end{table*}

In this section, we describe all state-of-the-art solutions of FL watermarking for IP protection. Note that all following papers are watermarking algorithms except the two last which are focusing on the ownership verification protocol.

\subsubsection{WAFFLE}
WAFFLE \cite{tekgul2021WAFFLE} is the first DNN Watermarking technique for FL. In this solution. The security hypothesis presented in the paper assumes that the server is a trusted party that embeds the watermark (\subscript{\textbf{S}}{\textbf{{1}}}) using a black-box watermarking technique using a trigger set. Clients cannot backdoor, poison, or embed their own watermarks since they are incentivized to maximize global accuracy. The adversary can only save the model and apply the post-processing technique as described in \ref{background}. Any trigger set that does not need any knowledge of the client's data can be used but the authors present a specific set that is more suitable for FL: the WAFFLEPattern. WAFFLEPattern is defined as a set of images containing random patterns with a noisy background. Basically, the server will embed the watermark into the global model using the two following functions:
\begin{itemize}
    \item \textbf{PreTrain}: Train an initialized model with the trigger set until the model has good accuracy on this set 
    \item \textbf{ReTrain}: Fine-tune the model with the trigger set until the model has learned the watermark 
\end{itemize}
\textbf{PreTrain} is used to embed the watermark in the model before the first round. During each round, after the aggregation process, the server uses \textbf{ReTrain} to re-embed the watermark into the model.

\subsubsection{FedIPR}
FedIPR \cite{li2022fedipr} is both a Black-Box and White-Box technique. This technique allows all clients to embed their own watermark in the global model (\subscript{\textbf{S}}{\textbf{{2}}}) without sharing secret information. Each watermark can be described as follow :
\begin{itemize}
    \item \textbf{Black-Box Watermark}: Each client generates a trigger set using the Projected Gradient Descent technique in a small Convolutional Neural Network (CNN) trained with the local data. 
    \item \textbf{White-Box Watermark}: Each client generates a random secret matrix and a location in the Batch-Normalisation layers to embed its message.
\end{itemize}
Both White-Box and Black-box watermarks are inserted using an additional loss during the local training. For \textbf{Black-Box Watermark}, the loss is exactly the same as the loss used for the main task but with a batch of the trigger set as input. For the \textbf{White-Box Watermark}, the loss used is a Hinge-like loss \cite{wahba1999support} between the original message and the reconstructed message. 
\subsubsection{FedTracker}
FedTracker \cite{shao2022fedtracker} is a watermarking technique that allows the server to embed a global Black-box watermark (\subscript{\textbf{S}}{\textbf{{1}}}) but also a White-box watermark specific to each client. The server is assumed to be a trusted party that has no access to natural data related to the primitive task. Some clients can be malicious and others are assumed to be honest. For malicious clients, they can copy and distribute the model but they follow the training process to maximize the global accuracy. Each watermark can be described as follow :
\begin{itemize}
    \item \textbf{Global Black-Box Watermark} : A trigger set is generated using the WAFFLEPattern method \cite{tekgul2021WAFFLE}.
    \item \textbf{White-Box Watermark}: The server generates a random secret matrix and a fingerprint for each client.
\end{itemize}
After the aggregation, the server embeds the \textbf{Global Black-Box Watermark} using the intuition of Continual Learning \cite{de2021continual} to avoid forgetting the main task. Then, for the \textbf{White-Box Watermark}, the loss used is a Hinge-like loss between the original message $b$ and the reconstructed message $\Tilde{b}$.

\subsubsection{Liu \textit{et al.} scheme} \cite{liu2021secure}
\label{liuetal}
The authors propose a client-side Black-box watermarking scheme (\subscript{\textbf{S}}{\textbf{{2}}}). This technique is designed to embed a watermark only from one client (which represents the initiator of the FL). This framework is designed to be used under HE since the server is not trusted. The authors assume also that other clients can be malicious. During the verification process, a fully-trustworthy third party is recruited to be the arbitrator. The initiator creates a trigger set composed of Gaussian noise images with a given label as a trigger set. Then, the client's model will overfit with this set like in \cite{adi2018turning}. To tackle the fact that this particular client will probably not be selected at each round, the authors introduce a scaling factor 
\begin{equation*}
  \lambda = \frac{N}{n}  
\end{equation*}
where $N$ is the number of clients and $n$ is the number of clients selected at each round. The client will then send its model weights multiplied by $\lambda$. According to the authors, this will be approximately equivalent to the case that this client is selected every iteration and the watermark will be embedded more easily.

\subsubsection{FedRight}
FedRight \cite{chen2023fedright} is a solution for the server to fingerprint the model in the FL framework (\subscript{\textbf{S}}{\textbf{{3}}}). DNN fingerprinting is a process in which instead of embedding a watermark in the model, we extract a fingerprint to identify this model \cite{cao2021ipguard}. The security assumptions presented by authors assume that the server is trusted and honest while clients may not be trustworthy. 
To do so, the server generates adversarial examples from a set of inputs (key samples). Then the server extracts the probability distribution of each prediction and feeds it to a detector with the key samples target. Then, during the verification process, this detector is used to predict whether the corresponding model is the good one or not.

\subsubsection{FedCIP}
FedCIP \cite{liang2023fedcip} is a White-Box watermarking framework that is compatible with FL security aggregation and allows tracking traitors. This solution allows the clients to watermark the FL model (\subscript{\textbf{S}}{\textbf{{2}}}). They assume that the server is honest but curious while model theft can be done by any malicious client. The server cannot directly modify the model's parameters. And the last security hypothesis relies on the fact that the trigger set should not rely on the original data. This technique relies on the concept of a unique watermark per client for a given number of rounds called a cycle. During a cycle, a portion of $cK$ of clients is designated to contribute during the following rounds. For each client, the watermark is unique during the cycle. If the round is the first iteration of the cycle, the algorithm will quickly replace the previous watermark with the new one. Otherwise, a small update is made to reinforce the watermark. The server divides the model parameters for all clients to avoid watermark conflicts and it records the participants at each round which is used for the verification process.

The verification is then performed using the "Federated Watermark" which is a concatenation of the watermark of all the contributors for a given cycle. If a client leaks the model multiple times during the FL, the authors point out that they can find the traitor by computing the intersection of the participants that have a high Watermark-Detection-Rate with the leak models. In particular, they can identify the traitor if $c^nK \leq 1$ where $n$ is the number of traitor leaks.

\subsubsection{Yang \textit{et al.} scheme \cite{yang2022watermarking}}
Yang \textit{et al.} \cite{yang2022watermarking} proposed a Black-Box watermarking framework based on Liu \textit{et al.} \cite{liu2021secure} solution (described in Section \ref{liuetal}) (\subscript{\textbf{S}}{\textbf{{2}}}). The main contribution of this paper is the proposed trigger set schema. Instead of using a classical Gaussian noise-based trigger set, they build images using permutation-based secret keys and noise-based patterns. The authors claim and demonstrate that this trigger set is resistant to ambiguity attacks. Whether the adversary tries to brute force the secret key or generate a new trigger set with its own key, both ways are hard to accomplish.

\subsubsection{Merkle-Sign}
Merkle-Sign \cite{li2021towards} is a framework focusing on ownership verification in a collaborative Clients-Server setting (\subscript{\textbf{S}}{\textbf{{3}}}). The authors propose a public verification protocol that uses the Merkle-tree \cite{becker2008merkle}. In this framework, the server use at each round an embedding function to insert two pieces of identity information (i.e. keys) into $M_G$: one that identifies the server and the other one the client that will receive the model. In parallel, the server uploads the tuple of keys and the tuple of verification function (which are generated by the watermarking embedding function) into a Merkle-tree with the recording time. In the final round, the server embeds also all keys generated by the clients into $M_G$ and update the Merkle-tree. This framework is also compatible in a Peer-to-Peer context. The associated Black-Box watermarking scheme relies on the training of an Auto-Encoder \cite{bank2020autoencoders} (AE) for each client. Then the server averages the received AE to obtain a final AE from which it can generate a trigger set using the keys as input for the decoder part. 

In the Merkle sign scheme, the server is trusted, and clients are considered honest but curious. The authors explore various attacks to assess the verification protocol's robustness, including ambiguity, spoiling, and traitor-tracing attacks. Ambiguity attacks can be countered by using an authorized time server or decentralized consensus protocol for timestamp authorization \cite{li2021secure}. Security against spoil attacks involves applying multiple watermarks to the model, allowing clients to prove ownership even if an adversary spoils one watermark. For traitor tracing, the server embeds unique watermarks for each client, enabling successful ownership declaration over pirated models. However, the paper lacks a protocol for identifying traitor clients after training has terminated, as the model is watermarked using information from all clients simultaneously.

\subsubsection{FedZKP}
The authors in \cite{yang2023fedzkp} propose a verification protocol based on the zero-knowledge proof, specifically on xLPN ZKP \cite{jain2012commitments} called FedZKP. FedZKP doesn’t require a trusted verifier to protect the confidentiality of the credentials, which reduces the risk of credential leakage. The security hypothesis presented in the paper assumes that clients are trusted entities, while the server and verifying authority are considered honest but curious (\subscript{\textbf{S}}{\textbf{{2}}}). Watermarking is conducted by the clients, each equipped with a public and private key. The hash of all clients' public keys is then embedded into the model using a white-box mode. The paper examines four attacks: fine-tuning, pruning, and targeted destruction attacks. The experimental results demonstrate the watermark's robustness against these attacks. The last attack discussed in the paper involves a scenario where a user is required to prove ownership of a DNN model by sharing their public key with the trusted authority. However, this opens up the possibility for an adversary who intercepts the key to falsely claim ownership of the model. To address this vulnerability, the verification authority implements a zero-knowledge proof protocol, ensuring that only the individual possessing the private key associated with the public key in the watermarking can successfully demonstrate ownership.

It is worth noting that while the majority of the proposed solutions focus on IP protection, some papers explore alternative approaches using FL watermarking. For instance, WMDefense \cite{zheng2022wmdefense} utilizes FL watermarking to detect malicious client updates by assessing the degree of watermark recession. In this section, we have covered solutions encompassing both Black-Box/White-Box and Client(s)/Server watermarking with various ownership verification schemes. The summary of these solutions is presented in Table \ref{tab:fl_watermarking}. Each solution is categorized based on the verification process (White-Box/Black-Box), the entity performing the watermarking (Client(s) or Server), compatibility with standard security tools, compatibility with communication efficiency systems such as client selection, and compatibility with poisoning defense mechanisms. 
Upon review of the table, it becomes apparent that many solutions have not been tested with security tools like Differential Privacy (DP) or Homomorphic Encryption (HE). Additionally, none of the solutions have undergone testing with Multi-Party Computation (MPC) and Trusted Execution Environments (TEE). The same observation applies to poisoning defense mechanisms, where comprehensive testing is yet to be explored across various solutions
\section{Discussion}
\label{discussion}
In this section, we identify and discuss specific challenges related to watermarking in FL. In particular, we evaluate how existing methods deal or not with these new challenges. 
\subsection{Trigger-set based-watermarking on the Server side}
\label{bbserver}
Black-Box watermarking consists in changing the behavior of the model. To do so, most methods let the model overfit on the trigger set by adding a regularization term in the loss function. Usual DNN watermarking techniques can easily be applied in \subscript{\textbf{S}}{\textbf{{2}}} and \subscript{\textbf{S}}{\textbf{{3}}}. However, it is not so easy in \subscript{\textbf{S}}{\textbf{{1}}}. When the client $C_k$ wants to watermark its model $M_{C_k}$, he can rely on two things : 

\begin{enumerate}
    \item Have access to its private dataset $D_{C_k}$
    \item Train the model on both the main task dataset $D_{C_k}$ and its trigger set $T_{C_k}$ at the same time
\end{enumerate}

A large number of Black-Box watermarking techniques need to build $T_{C_k}$ using $D_{C_k} = (X_i,Y_i)_{i=1}$ (as discussed in \ref{subsec:dnnwatermarking}). Using such techniques is motivated by the fact that training the model from these datasets is multi-task learning. Building $T_{C_k}$ from $D_{C_k}$ helps to reduce the negative impact on learning from two domains. Moreover, learning these two tasks together avoids catastrophic forgetting \cite{french1999catastrophic}.

In \subscript{\textbf{S}}{\textbf{{1}}}, since the server has not its own dataset, it cannot perform such type of watermarking. The choice is limited by using unrelated or noise-based inputs such as WafflePattern \cite{tekgul2021WAFFLE} or generating them from auto-encoders provided by clients such as Merkle-Sign \cite{li2021towards}.  As we said before, learning two tasks separately can be negative for Fidelity. FedTracker \cite{shao2022fedtracker} is the only server-side framework in which they use Continual Learning \cite{de2021continual} and a global gradient memory to not interfere with previously learned tasks.

In addition, this limitation leads to exposure to evasion attacks. During the Black-Box verification process, the owner will ask about the suspicious Application Programming Interface (API) that possibly contains his DNN. However, the attacker may evade this verification using a query detector \cite{hitaj2018have}. Since the trigger-set is built using images that are qualified to be Out-Of-Distribution (OOD), this implies an easier detection for the attacker \cite{yang2021generalized}. WAFFLE \cite{tekgul2021WAFFLE} authors confirm the intuition that the performances of such detectors depend a lot on the data quantity and capacities of the attacker.

\subsection{Aggregation functions}
\label{aggregation}
The most common aggregation function is \texttt{FedAvg} \cite{mcmahan2017communication} which consists of averaging clients' parameters after they perform multiple epochs on mini-batches. Each client weight matrix is multiplied by a scaling factor defined as $\frac{n_{C_k}}{n}$ where $n_{C_k}$ is the number of samples in $D_{C_k}$ and $n = \sum^K_k n_{C_k}$. Many aggregation functions emerged to meet various challenges in FL. Since the clients do not necessarily know which aggregation function the server is using, the proposed methods must be independent of this parameter.

For the Byzantine-attacks problem in which one or multiple clients try to disturb the FL process. These attacks can be simple noise weights or complex label-flipping backdoors. To leverage this problem, multiple aggregation functions appear to select only benign updates such as \texttt{Krum} \cite{blanchard2017machine} \texttt{Trim-mean} or \texttt{Bulyan} \cite{guerraoui2018hidden}. Since clients' watermarking techniques are sensitive to the embed message $b$ and the trigger set $T$, they keep their updates far from each other. A part of updates can be rejected for this reason if we use defensive aggregation techniques. As an example, FedIPR shows that the \textbf{White-Box Watermark} results are similar to \texttt{FedAvg} with a detection rate of 97.5\% using \texttt{Trim-mean}. However, the \textbf{Black-Box Watermark} reaches only 63.25\% of the watermark detection rate at the end of the FL process. Even if this score is enough to detect plagiarism, using a defensive aggregation function has a huge impact on the watermark. Liu \textit{et al.} \cite{liu2021secure} and its extension Yang \textit{et al.} \cite{yang2022watermarking} have not tested yet their solution with a defensive aggregation function but we can guess that multiplying weights by a so big scaling factor $\lambda$ can be easy to detect for \texttt{Krum} as a Byzantine attack as shown in similar example \cite{gu2021detecting}.

Another problem is that \texttt{FedAvg} performs well when the data are statistically homogeneously distributed among the clients. However, in real use cases, data are heterogeneous which may lead to difficulty for the model to converge using \texttt{FedAvg} \cite{li2020federated}. Existing watermarking techniques for FL have not evaluated methods that tackle this problem such as \texttt{FedProx} \cite{li2020federated}, \texttt{FedNova} \cite{wang2020tackling} or \texttt{SCAFFOLD} \cite{karimireddy2020scaffold}. If we want to use the proposed solution in a real Secure FL framework, these methods need to be tested in a such context which is actually not the case. 

\subsection{Clients selection}
For communication efficiency and when the number of clients is too high, a fixed number of clients is selected at each round. To do so, we simply randomly select $cK$ clients with $c \in \left] 0, 1 \right[$. This simple mechanism can have a big effect on the watermarking. In (\subscript{\textbf{S}}{\textbf{{1}}}), this effect should be insignificant, since the server embeds the watermark at each round regardless of $cK$. Unfortunately, the majority of papers have not tested this effect. On the other hand, (\subscript{\textbf{S}}{\textbf{{2}}}) is more able to be sensitive to the client selection process. Since each client wants to insert its own watermark, the watermark of not selected clients risks being degraded or removed in the global model. Authors of FedIPR \cite{li2022fedipr} show that for $c > 0.2$ the detection rate for both White-Box and Black-Box watermark is near 100\%. When $c \leq 0.2$ the detection rate associated with a feature-based watermark is still near 100$\%$. However, the detection rate for the backdoor falls to 62$\%$. FedCIP \cite{liang2023fedcip} framework is specially designed with the clients' selection routine and demonstrates good performances for several watermarking requirements.

\subsection{Cross-device setting}
All proposed papers are treating the case in which we have a small number of clients. The worse scenario is tested in Merkle-Sign in which 200 clients are enrolled in the federation. However, there is no solution that takes into account the cross-device setting. In this setting, a large number of clients (up to $10^{10}$ devices), are enrolled in the FL procedure \cite{kairouz2021advances}. These clients are not always reachable and they have a low dedicated computational power which is defined as a performance condition by the authors of WAFFLE.

In the Black-Box setting, the problem can come from the low computation power that does not allow the client to perform more computations to increase the batch size using trigger-set methods. For the White-Box setting, the bottleneck would be the \textbf{Capacity} as mentioned in Section \ref{def:wfl}. In particular in cases where the proposed methods are tested using Normalization layers such as FedIPR or FedTracker which limits the overall embedding capacity. And it leads to difficulty for each client to embed its vector $b$ without conflicting in the face of other clients' watermarks.

\subsection{Differential Privacy and Homomorphic Encryption}
\label{differential} 

Since Federated Learning (FL) ensures the privacy of clients' data by sharing the model or gradient updates between the server and clients (or directly among clients), there are concerns about potential attacks, such as membership inference, which can reveal the presence of specific data points \cite{melis2019exploiting}. To address this issue, Differential Privacy (DP) is often employed, providing robust privacy guarantees for algorithms working on aggregate databases \cite{abadi2016deep, wang2023ppefl}. In FL, a common DP technique involves introducing Gaussian random noise to the gradients sent to the aggregator, adding an extra layer of privacy protection.

Another cryptographic measure to enhance security in Client-Server FL is Homomorphic Encryption (HE) \cite{bellafqira2019secure, ma2022privacy, jin2023fedml}. In this approach, clients can protect their model updates using their public keys, encrypting the data before sending it to the server. The server performs the model aggregation (usually using \texttt{FedAvg}) in the encrypted space, ensuring that the server cannot misuse the privilege of having access to individual client updates to learn about private datasets. Clients can then use their private keys to decrypt the aggregated global model once it is received.

The combination of Differential Privacy and Homomorphic Encryption addresses privacy and security concerns in different aspects of FL. While DP protects against attacks attempting to extract sensitive information from gradient updates, HE safeguards the clients' data during aggregation, preventing unauthorized access by the server. In the context of watermarking, the use of HE is particularly suitable for scenario (\subscript{\textbf{S}}{\textbf{{2}}}), as clients can access their model parameters to embed watermarks securely. However, a challenge remains for scenario (\subscript{\textbf{S}}{\textbf{{1}}}), as the server cannot decrypt the model parameters to perform watermarking embedding.

\subsection{Watermarking for Non-Client-Server framework}

Decentralized FL (\subscript{\textbf{S}}{\textbf{{2}}}) is an interesting framework in which clients do not need a server to perform the model aggregation. The proposed methods seem to be applicable to this setting since watermarking the model from the client side does not require the server. However, Merkle-Sign is a unique solution that extends to the decentralized setting. We can also cite Split-Learning in which the server has a part of the network and clients have another part. Performing White-Box watermarking as in \cite{uchida2017embedding} can be more difficult for the server or for clients. In both cases, they have access to a part of the model parameters that can be arbitrarily small.

In the U-shape Split Learning architecture, the server has only the middle part of the model and the clients have the first and last layers. In this setting, performing a BlackBox watermarking on the server side is hard since it cannot use its inputs and labels on the model.

\subsection{Attacks from clients and server}

When we analyze  (\subscript{\textbf{S}}{\textbf{{1}}}) and  (\subscript{\textbf{S}}{\textbf{{2}}}) scenarios, we can see that each one has different parameters to play with whether for watermarking or disrupt it. The server can control the selected participants or how to aggregate the model parameters. It has also sometimes a clear view of clients' parameters at each round. However, it does not have data and it cannot fully control if clients are strictly following the training process. On the other hand, clients have their private dataset and they can send the update that they want. Nevertheless, they have no control over what happens with their updates at the server level.

If the server wants to avoid a subset of clients to watermark the model, it can use methods proposed for Byzantine attacks \cite{shi2022challenges} detection. In particular, attacks that consist of multiplying the weights by a scaling factor to replace or have a bigger impact in the global model are easy to detect \cite{gu2021detecting}. The proposed method by Yang \textit{et al.} \cite{liu2021secure}, without HE, is then easily removable and the global model will not be watermarked. A solution to catch backdoored models was presented in \cite{tolpegin2020data} \cite{xi2021batfl}. Then all proposed solutions that rely on a backdoor-based watermarking can be rejected.

Clients can also try to disturb the watermarking process. FedIPR (\subscript{\textbf{S}}{\textbf{{2}}}) \cite{li2022fedipr} authors present the free-riders problem in which some clients do not contribute to the training of $M_G$ and the watermarking process. Even if with their solution, they have no important impact on the watermarking, no testing has yet been done on (\subscript{\textbf{S}}{\textbf{{1}}}). Another attack that is specific to FL as described in FedRight \cite{chen2023fedright} and WAFFLE \cite{tekgul2021WAFFLE} consists in the fact that multiple clients will use their models and private dataset. As mentioned in Section \ref{bbserver}, an evasion attack works better when using multiple datasets are used to train the detector. But it is also possible to fine-tune the model with the combined dataset.

In summary, none of the client-side solutions have undergone testing with poisoning detection mechanisms such as anomaly detection or defensive aggregation functions like \texttt{Krum} or \texttt{GeoMed}. As for server-side solutions, no watermarking algorithm has been found compatible with cryptographic tools like Homomorphic Encryption (HE). Additionally, certain aggregation functions are not compatible with FL watermarking in both cases. Concerning data repartition, only a small portion of the proposed solutions have been tested in a non-I.I.D scenario, with no solutions tested in vertical data repartition or split learning. Despite the multitude of proposed solutions for FL watermarking, no FL framework (as presented in Table \ref{tab:fl_frameworks}) has integrated a tool to embed a watermark.

\section{Conclusion}
\label{sec:conclusion}

In the context of Federated Learning (FL), watermarking has gained significant interest due to the limitations of applying classical Deep Neural Network (DNN) watermarking in a collaborative setting. Addressing data distribution, new distributed threat models, and incorporating additional security mechanisms becomes crucial when designing an efficient solution for collaborative ML watermarking. This survey paper provides a comprehensive overview of FL watermarking, including its definition, requirements, and a taxonomy of existing methods, along with their security assumptions. We have also discussed the constraints that watermarking in the FL context can encounter concerning privacy-preserving issues. It is our hope that this paper will contribute to the advancement of research in the area of Intellectual Property (IP) protection, particularly in the context of vertical learning and split learning. Furthermore, we anticipate that these solutions will be seamlessly integrated into secure FL frameworks, enabling their deployment across a wider range of use cases.

\bibliographystyle{IEEEtran} 
\bibliography{bibliography} 

\end{document}